# Breakdown of Fermi degeneracy in the simplest liquid metal

M. Zaghoo[1]*, T. R. Boehly[1], J. R. Rygg[1], P. M. Celliers[2], S. X. Hu[1], and G. W. Collins[1]

[1]Laboratory for Laser Energetics, University of Rochester, Rochester, NY 14623-1299, USA

[2]Lawrence Livermore National Laboratory, Livermore, CA 94550-9234, USA

*Correspondence to: mzag@lle.rochester.edu

**We are reporting the observation of the breakdown of electrons' degeneracy and emergence of classical statistics in the simplest element: metallic deuterium. We have studied the optical reflectance, shock velocity and temperature of dynamically compressed liquid deuterium up to its Fermi temperature. Above the insulator–metal transition, the optical reflectance shows the distinctive temperature-independent resistivity saturation, which is prescribed by Mott's minimum metallic limit, in agreement with previous experiments. At $T > 0.4\ T_F$, however, the reflectance of metallic deuterium starts to rise with a temperature-dependent slope, consistent with the breakdown of the Fermi surface. The experimentally inferred collisional time in this region exhibits the characteristic temperature dependence expected for a classical Landau-Spitzer plasma. Our observation of electron degeneracy lifting extends studies of degeneracy to new fermionic species, electron Fermi systems and offers an invaluable benchmark for quantum statistical models of Coulomb systems over a wide range of temperatures relevant to dense astrophysical objects and ignition physics.**

The phase diagram of Fermi matter can be demarcated by regions where quantum or classical effects are uniquely dominant. In dense quantum plasmas, like degenerate electrons in metals, the constitutive particle statistics underpinning the thermodynamic and transport properties as well as the strength of the interaction differ significantly from classical ideal plasmas *(1-5)*. The two key energy scales that dictate the relevant thermodynamic statistics and the strength of the interparticle interactions are the Fermi energy and the Coulomb energy, respectively *(6,7)*. The ratio of the thermal to the Fermi energy defines the degree of quantum degeneracy, $\Theta = T/T_F$, and at increasing densities, $T_F = (\hbar^2/2m)(3\pi^2 n)^{2/3}$, rises as $n^{2/3}$, where $n$ is the carrier density, m is the fermionic mass. In the limit of sufficiently high densities such as metals, matter is Fermi degenerate ($\Theta \ll 1$), whereas dilute or sufficiently hot plasmas ($\Theta \gg 1$) remain classical. On the other hand, the ratio of the Coulomb potential to the kinetic energy defines the dimensionless Coulomb coupling parameter $\Gamma = e^2/ak_BT$, where $a = (3/4\pi n)^{1/3}$ is the ion sphere radius, $e$ is the electron charge, and $k_B$ is the Boltzmann constant. Strongly ion-coupled plasmas with $\Gamma \gg 1$ exhibit long-range liquid-like spatial correlations (*6*) that eventually submit to Wigner crystallization at high enough $\Gamma$ (*1*). The weak coupling regime is characterized by $\Gamma \ll 1$ and the plasma behaves classically as an ideal gas where small-angle scattering describes the collisions.

An accurate description of the intermediate regime between these two extremes remains an outstanding challenge in statistical physics (*7–9*). A multitude of theoretical efforts attempt to characterize the thermodynamic state of model systems such as the free electron gas or one-component plasmas across the Pressure-Temperature space. In dense plasmas, the quantum-to-classical crossover occurs in the warm-dense-matter regime, where a complex interplay of Coulombic, thermal, and quantum effects characterize systems that are too hot for cold Fermi-liquid condensed-matter frameworks, yet too dense for standard classical plasma theories (*6,7*). Fluid metallic deuterium represents a unique state to investigate some of the aforementioned effects. Unlike all other metals, it has no bound electrons, and at sufficiently high density, it constitutes the archetypal one-component Coulomb system: where the ions are weakly coupled to the nearly-free conduction electrons (6,11).

Thus far, experimental studies probing the emergence of quantum statistical effects from classical regimes have been limited to strongly coupled low-temperature systems such as dilute solutions of liquid $^3$He (*12*) and magnetically or optically trapped ultracold alkali gases (*13,14*). Attempts to experimentally study the degeneracy effects in denser systems are stymied by the difficulty of confining these systems at sufficiently high temperatures while maintaining thermodynamic equilibrium (*15*). Here, we report the observation of the crossover of dense deuterium from a degenerate strongly coupled metal to a classical weakly coupled plasma. In strongly shocked deuterium along the Hugoniot, corresponding to conditions well above the insulator–metal transition, the density is nearly constant with increasing shock pressure (*16-18*). This allowed us to isolate the effect of the recorded temperature, and consequently degeneracy, on the electron-transport properties of the system. These properties are deduced from the optical reflectivity of the shocks.

Spherical deuterated carbon (CD) shells filled with liquid deuterium were directly irradiated with several 100-ps laser pulses to produce a spherically convergent shock wave in liquid deuterium ($\rho_0 = 0.172$ g/cm$^3$). The laser pulse launched an impulse drive that produces an initially strong (up to ~5.5 Mbar) but unsupported shock that decays in pressure, and therefore shock velocity, as it propagates. Velocity interferometer system for any reflector (VISAR) and a streaked optical pyrometer (SOP) were used to measure temporally resolved shock velocity and self-emission profiles, respectively, for these shocks propagating in liquid deuterium. Figure 1(A) shows a typical VISAR record, which is a 1-D image (vertical axis) of the shock that is streaked in time (horizontal axis). Imposed on that image is a series of fringes whose vertical position (fringe phase) is proportional to the shock velocity. Details of the experimental geometry are described elsewhere (19).

The shock velocity versus time in Fig 1(A) reveals the decay rate is sufficiently slow compared to the equilibration time, see SM. The optical diagnostics probe the shock front to a depth of about 30 to 40 nm, the optical skin depth, which is deep enough to sample an equilibrated plasma state, yet shallow enough to track the varying shock state continuously during the shock decay. The absolute reflectance (*R*) was inferred from the time-resolved VISAR laser intensity reflected from shocked deuterium, normalized to the stationary and unshocked gold cone. The temperature was determined by measuring the spectral radiance from the shock front, calibrated against a NIST-traceable tungsten filament (*20*). This spectral radiance was converted to temperature assuming the shock front is a gray-body emitter with a wavelength-independent emissivity (ε). The velocity-dependent ε was determined from the optical reflectance at 532 nm, (ε = 1–R).

In this study, we observed shocks that decay from 60 km/s to 35 km/s, which corresponds to ~5.5 to ~0.5 Mbar in pressure. Over this range, the density is nearly constant (ρ=0.774, $T_F$=13.8 eV), but the temperature ranges from 3 to 11 eV (1 eV is 11,603 K). Pressure is estimated from P=ρUsUp, where Us is the shock velocity, Up is the particle velocity, and the constant density value ρ could vary from 4.2-4.5 (See SM). Consequently, we are able to probe the electronic properties (as deduced from its optical reflectivity) of deuterium compressed to 0.774 g/cm$^3$ as a function of temperature. At the lower pressures, the shocked deuterium is strongly coupled and highly degenerate (Γ >> 1, Θ << 1); as the temperature increases, both of these defining characteristics change primarily. The present data exhibit two behaviors as these parameters change. For 0.15 < Θ < 0.4 and 2.6 < Γ < 6, we observe a constant optical reflectance (~40%) consistent with reflectivity saturation observed in previous shock-wave experiments [see Fig. 2] (*16-18*). This value is described by the Mott–Ioffe–Regel (MIR) minimum metallic conductivity limit, where the electron–ion relaxation time is set by the interatomic spacing *a* and the Fermi velocity $v_F$ through $\tau_{min} = a/v_F$. An equivalent derivation of this minimum could be deduced from the elementary quantum picture of conduction, which maintains that the minimum mean free path that a quasiparticle can travel is set by the Heisenberg uncertainty principle through $\Delta x k_F \sim 1$ (*21,22*), where $k_F$ is the Fermi wave vector. We also show the results of previous experiments on shocked deuterium in comparison to our data. In those experiments the deuterium transitions from an insulator to a conductor and then sits at the "saturation" value of ~ 40% as Θ increases and Γ decreases. At full ionization, the MIR limit predicts a minimum optical reflectivity of 0.38 for 532-nm light, in excellent agreement with our data as well as previous data.

The second behavior occurs above Θ > 0.4 (T ~ 5 eV), where the reflectance rises continuously to ~0.7 at T ~ 11 eV, as shown in Fig. 2. Over this region the Coloumb coupling relaxes as Γ approaches 1 and the fluid is now moderately coupled. At 5 eV, the deuterium is expected to be fully ionized; the change in the reflectance is consequently the result of a change in the scattering time. To test this, we investigated whether a partial ionization rise could account for the observed increase in reflectance at higher temperatures. In the SM we show that, for a fixed relaxation time, 0.7 to full ionization increases the reflectivity to ~40%, well short of the observed 70%. Similarly, a different range of densities than the one considered here cannot account for the observed increase in reflectance.

To clearly elucidate the effect of the scattering times $\tau$ on the observed reflectance, we determined $\tau$ for the data recorded using the standard Fresnel relations and the free-electron model (see SM). For a specularly smooth interface, the measured reflectance at a given frequency is $R(\omega) = |(N_{D2} - N_{MetD})/(N_{D2} + N_{MetD})|^2$ where $N_{D2}$ is the index of refraction of unshocked molecular deuterium layer and $N_{MetD}$ is the complex index of refraction of shocked deuterium. In the free electron model $N_{metD}^2 = 1 - \omega_P^2/(\omega^2 + j\omega/\tau)$ where $\omega_P$ is the plasma frequency, which is directly related to the carrier density n by $\omega_p^2 = 4\pi n e^2/m_e$ and $\tau$ is the electron relaxation time. We have fitted our measured reflectance at 532 nm to the relation above to determine the relaxation time, assuming full ionization (a plasma frequency of 17.9 eV).

Figure 3 shows that in the highly degenerate regime, the MIR limit with the carriers traveling at $v_F$ describes the data. This restriction in the velocity of the degenerate conducting carriers is a direct manifestation of the Pauli-exclusion principle that permits the formation of the Fermi sphere. Our data, therefore, suggests that up to $T/T_F \sim 0.4$, a well-defined Fermi surface still exists in the metallic fluid. Above this temperature, however, the increase in the inferred relaxation times implies that the restriction in the allowed velocities is no longer valid and longer relaxation times (higher thermal velocities) are needed to account for the increased reflectance. A fit to the relaxation time in this region reveals that $\tau \sim T^{1.55 \pm 0.04}$ (see Fig. 3). This is strikingly close to the classical nondegenerate limit expected for ideal plasmas, $\tau \sim T^{1.5}$. In standard plasma theories, such dependence arises because the carriers assume thermal velocities (*4*) or $k_B T = (1/2)mv^2$. In Fig. 3(A) we compare the experimentally inferred electrical conductivities values with those predicted by two of the commonly used transport models in dense plasmas (*23,24*), see SM for details of the calculations. Although the analytical models asymptotically converge to the two limits—degenerate Ziman and nondegenerate Spitzer—they don't capture the location of the crossover nor the magnitude of the intermediate regimes. This crossover has striking implications for the thermodynamic, electron, and mass transport properties of the dense conducting fluid. For example, the sign of the system's chemical potential $\mu(T)$ changes from positive in Fermi–Dirac limit to negative in Maxwellian plasmas, while the heat capacity, $C_v$, goes from $C_v \propto T/T_f$ in the degenerate limit to $C_v \sim 3R$. Similarly, the transport coefficients assume a different temperature dependence across these two thermodynamic regimes: thermal conductivity $k \propto T$ in the degenerate regime, but $k \propto T^{5/2}$ in the ideal plasma limit.

It is instructive to compare our reported crossover between the classical and the degenerate limits in a dense deuterium electronic system to previous observations carried out in dilute liquid $^3$He or ultracold alkali gases. In those systems, a similar crossover in the temperature dependence of dynamical properties of an atomic fermionic system signals the onset of the quantum statistics (see comparison in Fig. 4 and SM). Despite the 8 to 12 orders of magnitude difference in absolute temperature and density, the physics of degeneracy lifting in Fermi systems remains the same. Our results are consistent with path-integral Monte Carlo calculations of dense hydrogen plasmas (*25*), which revealed significant permutation/exchange of the electrons in the plasma occurring at $T < 0.4\ T_F$ for different densities. Above this temperature, the probability of quantum exchange between two or more electrons was found to rapidly decline (*25*). Since permuting/exchanging electrons are a requisite for the formation of a Fermi surface, at an increasing temperature, the electrons are no longer degenerate and the Fermi sphere breaks down, in excellent accord with our observation.

We have studied the optical reflectivity of dense metallic deuterium over a range of temperatures spanning almost an order of magnitude. The observed saturation of optical conduction up to $T \sim 5$ eV is consistent with the MIR limit, which prescribes a temperature-independent relaxation time. Resistivity saturation is a phenomenological hallmark of Fermi-liquid behavior in metals (*21*), indicating the persistence of elementary quasiparticles in our metallic system up to $0.4\ T_F$. This was previously suggested in *ab initio* calculations, which found that the quasiparticle peak, despite broadening, remains well defined in a jellium system up to few $0.1 \times T_F$ (*26*). More sophisticated dense plasma models employing the Fermi golden rule in transport studies finds similar results (*27*). The marked rise in reflectance above this condition requires a temperature-dependent relaxation time that is well described by the classical Spitzer–Landau limit for nondegenerate plasmas. To the best of our knowledge, our observation provides the first experimental demonstration for this crossover in a dense electron system, extending studies of degeneracy to fluid metals or dense plasmas, and new regimes of weak interionic couplings. The onset of this crossover is different from what is coarsely described in the general literature of dense plasmas, but consistent with more sophisticated dense plasma transport models. Our result should allow a predictive criterion for the degeneracy condition occurring in compact astrophysical bodies and inertial confinement fusion targets. In the former, this condition is often invoked to delineate the boundary between the atmosphere and the degenerate core (*28*), while in the latter it serves as a crucial guide for the desired range of temperatures at which the nuclear fuel should be maintained during implosion (*6,7*). This should provide a new impetus for the search for quantum phenomena in warm dense matter, where a more-complex structure than the one revealed here might separate the Fermi-liquid and classical regimes (*9*).

**Fig 1.**

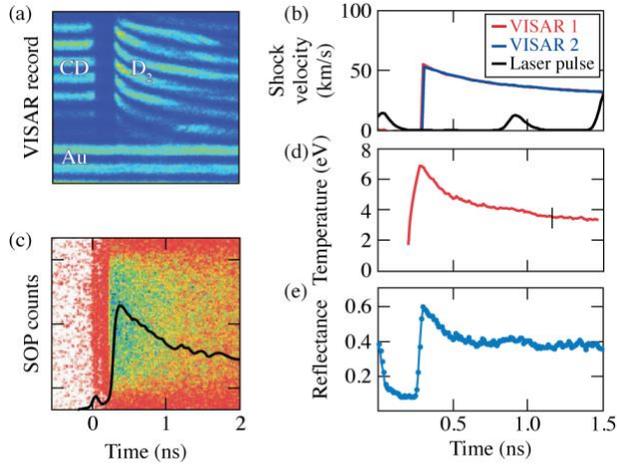

**Fig 2.**

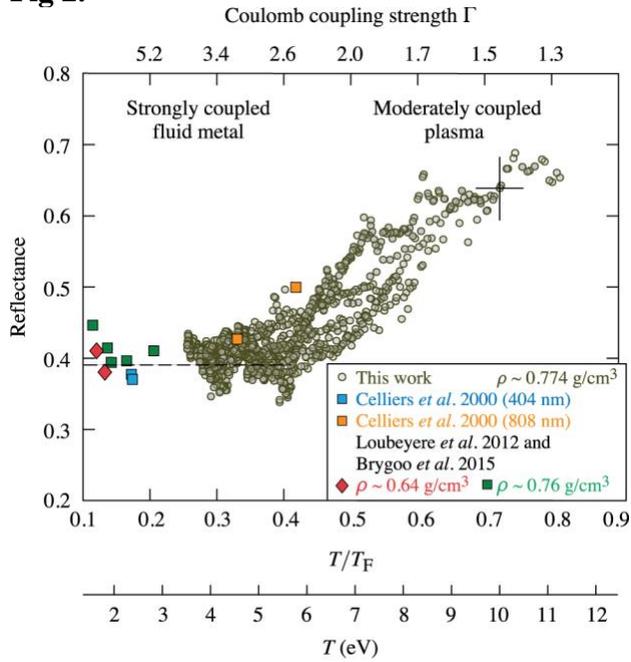

**Fig 3.**

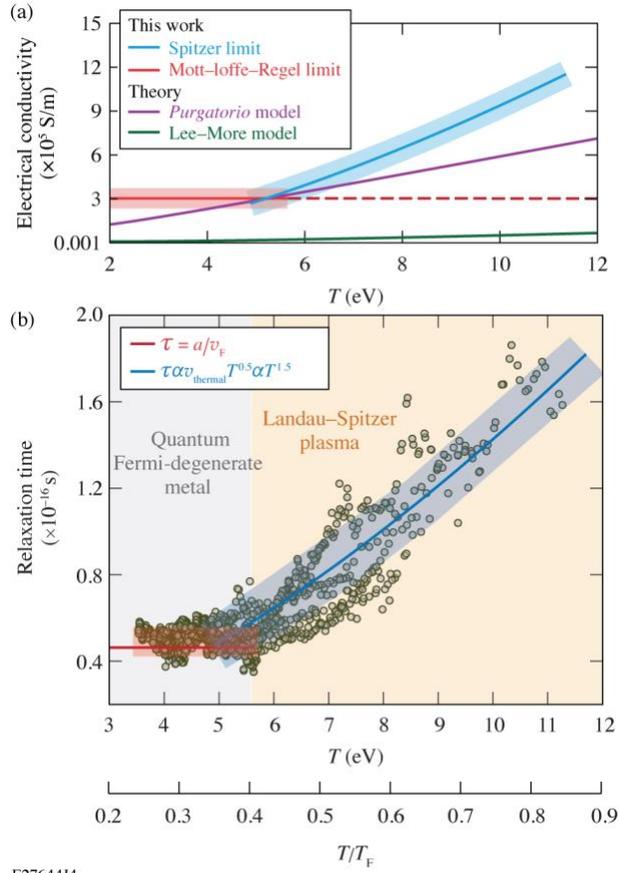

**Fig 4.**

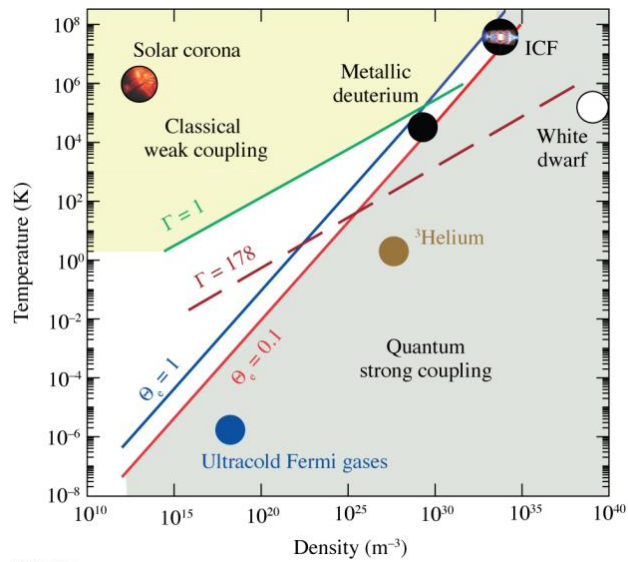

**References and Notes:**

1. D. M. Ceperley, B. J. Alder, Ground state of the electron gas by a stochastic method. *Phys. Rev. Lett.* **45**, 566–569 (1980).

2. M. N. Rosenbluth, W. M. MacDonald, D. L. Judd, Fokker–Planck equation for an inverse-square force. *Phys. Rev.* **107**, 1–6 (1957).

3. L. Spitzer, Jr., R. Härm, Transport phenomena in a completely ionized gas. *Phys. Rev.* **89**, 977–981 (1953).

4. L. D. Landau, The transport equation in the case of Coulomb interactions, *Phys. Z. Sowjetunion* **10**, 154–164 (1936).

5. S. Ichimaru, Strongly couple plasmas: high-density classical plasmas and degenerate electron liquids. *Rev. Mod. Phys.* **54**, 1017–1059 (1982).

6. R. P. Drake, "High-energy-density physics: Fundamentals, inertial fusion, and experimental astrophysics," in Shock Wave and High Pressure Phenomena (Springer, Berlin, 2006).

7. D. Kremp, M. Schlanges, and W.-D. Kraeft, Quantum Statistics of Nonideal Plasmas (Springer, Berlin Heidelberg, 2005).

8. T. Dornheim et al., Ab Initio Quantum Monte Carlo Simulation of the Warm Dense Electron Gas in the Thermodynamic Limit. Phys. Rev. Lett. 117, 156403 (2016)

9. J. Daligault, Crossover from classical to Fermi liquid in dense plasmas. Phys. Rev. Lett. 119, 045002 (2017).

10. S. Groth et al., *Ab initio* Exchange correlation Free Energy of the Uniform Electron gas at Warm Dense Matter conditions. Phys. Rev. Lett. **119**, 135001 (2017)

11. M. Zaghoo, I. F. Silvera, Conductivity and dissociation in metallic hydrogen implications for planetary interiors. *Proc. Natl. Acad. Sci.* **114**, 11873 (2017).

12. W. R. Abel, R. T. Johnson, J. C. Wheatley, W. Zimmermann, Thermal conductivity of pure He$^3$ and of dilute solutions of He$^3$ in He$^4$ at low temperatures. *Phys. Rev. Lett.* **18**, 737 (1967).

13. B. DeMarco, D. S. Jin, Onset of Fermi degeneracy in a trapped atomic gas. *Science* **285**, 1703 (1999).

14. I. Bloch, J. Dalibard, Wilhelm Zwerger, Many-body physics with ultracold gases. *Rev. Mod. Phys.* **80**, 885–964 (2008).

15. H. M. Milchberg, R. R. Freeman, S. C. Davey, R. M. More, Resistivity of a simple metal from room temperatures to $10^6$K. *Phys. Rev. Lett.* **61**, 2364 (1988).

16. P. M. Celliers *et al.*, Shocked-induced transformation of liquid deuterium into a metallic fluid. *Phys. Rev. Lett.* **84**, 5564 (2000).

17. M. D. Knudson *et al.*, Equation of state measurements in liquid deuterium to 70 GPa. *Phys. Rev. Lett.* **87**, 225501 (2001).

18. P. Loubeyre *et al.*, Extended data set for the equation of state of warm dense hydrogen isotopes. *Phys. Rev. B* **86**, 144115 (2012).

19. T. R. Boehly *et al.*, Velocity and timing of multiple spherically converging shock waves in liquid deuterium. *Phys. Rev. Lett.* **106**, 195005 (2011).

20. J. E. Miller *et al.*, Streaked optical pyrometer system for laser-driven shock-wave experiments on OMEGA. *Rev. Sci. Instrum.* **78**, 034903 (2007).

21. N. E. Hussey, K. Takenaka, H. Takagi, Universality of the Mott–Ioffe–Regel limit in metals. *Phil. Mag.* **84**, 2847–2864 (2004).

22. N. F. Mott, Conduction in non-crystalline systems IX. The minimum metallic conductivity. *Philos. Mag.* **26**, 1015–1026 (1972).

23. Y. T. Lee, R. M. More, An electron conductivity model of dense plasmas. *Phys. Fluids* **27**, 1273–1286 (1984).

24. P. A. Sterne, S. B. Hansen, B. G. Wilson, W. A. Isaacs, Equation of state, occupation probabilities and conductivities in the average atom Purgatorio code. *High Energy Density Phys.* **3**, 278–282 (2007).

25. C. Pierleoni, D. M. Ceperley, B. Bernu, W. R. Magro, Equation of state of hydrogen plasma by path integral Monte Carlo simulation. *Phys. Rev. Lett.* **73**, 2145–2149 (1994).

26. L. X. Benedict, C. D. Spataru, S. G. Louie, Quasiparticle properties of a simple metal at high electron temperatures. *Phys. Rev. B* **66**, 085116 (2002).

27. J. Daligault & G. Dimonte, Correlation effects on the temperature-relaxation rates in dense plasmas. Phys. Rev. E. 79, 056403 (2009).

28. H. M. Van Horn, Dense astrophysical plasmas. Science 252, 384–389 (1991).

29. D. G. Hicks et al., Laser-driven shock compression of fluid deuterium from 45 to 220 GPa. Phys. Rev. B 79, 014112 (2009); 85, 099901(E) (2012).

30. T. R. Boehly *et al.*, Demonstration of the shock-timing technique for ignition targets on the National Ignition Facility. *Phys. Plasmas* **16**, 056302 (2009).

31. P. M. Celliers *et al.*, Line-imaging velocimeter for shock diagnostics at the OMEGA Laser Facility. *Rev. Sci. Instrum.* **75**, 4916–4929 (2004).

32. J. M. McMahon, M. A. Morales, C. Pierleoni, D. M. Ceperley, The properties of hydrogen and helium under extreme conditions. *Rev. Mod. Phys.* **84**, 1607–1653 (2012).

33. M. Zaghoo, R. Husband, I. F. Silvera, Striking Isotope Effect on the Metallization Phase Lines of Liquid Hydrogen and Deuterium. *Phys. Rev. B* **98**, 104102 (2018).

34. M. Zaghoo, Dynamic conductivity and partial ionization in dense fluid hydrogen. *Phys. Rev. E* **97**, 043205 (2018).

35. W. R. Magro, B. Militzer, D. Ceperley, B. Bernu, C. Pierleoni, "Restricted Path Integral Monte Carlo Calculations of Hot, Dense Hydrogenin" in *Strongly Coupled Coulomb Systems,* G. J. Kalman, J. M. Rommel, K. Blagoev, Eds. (Springer, Boston, MA, 2002), pp. 337–340.

36. B. Wilson, V. Sonnad, P. Sterne, W. Isaacs, Purgatorio – A new implementation of the Inferno algorithm. *J. Quant. Spectrosc. Radiat. Transf.* **99**, 658–679 (2006).

37. L. Caillabet, S. Mazevet, P. Loubeyre, Multiphase equation of state of hydrogen from *ab initio* calculations in the range 0.2 to 5 g/cc up to 10 eV. *Phys. Rev. B* **83**, 094101 (2011).

38. N. M. Tubman, E. Liberatore, C. Pierleoni, M. Holzmann, D. M. Ceperley, Molecular-atomic transition along the deuterium Hugoniot curve with coupled electron-ion Monte Carlo simulations. *Phys. Rev. Lett.* **115**, 045301 (2015).

**Figure Captions**

**Fig. 1.** (**A**) Example of the raw velocity interferometer system for any reflector (VISAR) data record from a typical experiment with decaying shock velocity. The fringe phase determines the apparent velocity of the shock, while its amplitude determines the reflectance of the shock front.. Before $t = 0$, there are no fringe movements; at $t = 0$ the signal disappears as a result of the photoionization of the CD shell and then breaks out in the deuterium samples. The shock decelerates in the sample since it is unsupported. (**B**) Streaked optical pyrometer (SOP) data

showing the corresponding thermal self-emission from the target on the same time axis. (**C**) Shock-velocity time profile analyzed from the VISAR record. (**D**) The temperature of the sample determined from the self-emission SOP data over a band centered at 650 nm. (**E**) A record of the reflectance for the same data set.

**Fig. 2.** The reflectance of the deuterium plasmas as a function of the degeneracy $\Theta$ and Coulomb coupling $\Gamma$ parameters in comparison with previous shock-wave experiments. The data combine ten different experiments, showing great consistency. All the data shown are for 532-nm light, except for those of Celliers *et al.* were collected at 404- and 808-nm light. The Fermi temperature, and the ion coupling parameter $\Gamma$ were calculated for a deuterium density of 0.774 g/cm$^3$, assuming full ionization. The error bars on a datum are representative of the systematic uncertainty in both the temperature and the reflectance. The dashed line corresponds to the expected reflectance at full ionization in the Mott–Ioffe–Regel (MIR) Fermi limit.

**Fig. 3.** (**B**) Relaxation times of the deuterium plasmas as a function of degeneracy parameter. The light grey region designates the strong-scattering Mott regime where the relaxation time is described by the ratio of the minimum atomic spacing to the Fermi velocity. This ratio is shown by the solid red line and equals $4.574 \times 10^{-17}$s. In the light yellow region, the increase in the determined relaxation time is well described by the Spitzer classical limit $\tau \sim T^{1.5}$. The solid blue line shows the best fit of the relaxation time to temperature, which reveals $\tau \sim T^{1.552 \pm 0.04}$ with 95% confidence. The shaded regions show the $\pm 1$-$\sigma$ bounds to the data in the two different regimes. The crossover in the carriers' velocities from the Fermi limit to the thermal Maxwellian limit evinces the breakdown of Fermi degeneracy in our metallic deuterium. (**A**) The calculated

electrical conductivity of deuterium plasma as a function of temperature is shown for different models: The MIR and Spitzer limits are determined from this work, whereas the green and purple lines are from two of the commonly employed plasma transport models, Lee–More and Purgatorio respectively.

**Fig. 4.** Phase diagram of matter showing the parameter space for various Fermi systems in a log–log plot of temperature versus carrier number density. The solid lines for the $\Theta$ parameter demarcate the regions of high and low degeneracy for electrons; different lines are expected for $^3$He and ultracold gases, depending on their particle masses (see SM). $\Gamma = (e^2/ak_BT) = 1$ roughly separates strong and weak coupling, while $\Gamma = 178$ delineates the condition for Wigner crystallization. Our present work on the collapse of degeneracy in metallic deuterium at increasing $T$ is contrasted to previous experiments probing the emergence of degeneracy in cold dilute systems at decreasing $T$ close to their respective $T_F$.

## Acknowledgment

This material is based upon work supported by the Department of Energy National Nuclear Security Administration under Award Number DE-NA0001944, the University of Rochester, and the New York State Energy Research and Development Authority. We thank Jerome Daligault, and the anonymous reviewers for constructive comments. MZ acknowledges insightful discussions with A. Omran.

This report was prepared as an account of work sponsored by an agency of the U.S. Government. Neither the U.S. Government nor any agency thereof, nor any of their employees, makes any warranty, express or implied, or assumes any legal liability or responsibility for the accuracy, completeness, or usefulness of any information, apparatus, product, or process disclosed, or represents that its use would not infringe privately owned rights. Reference herein to any specific commercial product, process, or service by trade name, trademark, manufacturer, or otherwise does not necessarily constitute or imply its endorsement, recommendation, or favoring by the U.S. Government or any agency thereof. The views and opinions of authors expressed herein do not necessarily state or reflect those of the U.S. Government or any agency thereof.